\documentclass[12pt,preprint]{aastex}

\slugcomment{}

\shorttitle{}
\shortauthors{}

\begin{document}

\title{FIRST RESULTS FROM SPARO: EVIDENCE FOR LARGE-SCALE TOROIDAL
MAGNETIC FIELDS IN THE GALACTIC CENTER}

\author{G. Novak,\altaffilmark{1} D. T. Chuss,\altaffilmark{1} 
T. Renbarger,\altaffilmark{1} G. S. Griffin,\altaffilmark{2} 
M. G. Newcomb,\altaffilmark{2} J. B. Peterson,\altaffilmark{2} 
R. F. Loewenstein,\altaffilmark{3} D. Pernic,\altaffilmark{3} 
and J. L. Dotson\altaffilmark{4}}


\altaffiltext{1} {Department of Physics and Astronomy, Northwestern 
University, Evanston, IL 60208, \mbox{g-novak@northwestern.edu}}

\altaffiltext{2} {Department of Physics, Carnegie Mellon University, 
Pittsburgh, PA 15213}

\altaffiltext{3} {Yerkes Observatory, University of Chicago, 
Williams Bay, WI 53191}

\altaffiltext{4} {NASA/Ames Research Center, MS 245-6, Moffett Field, 
CA 94035}

\begin{abstract}

We have observed the linear polarization of 450 \micron\ continuum
emission from the Galactic center, using a new polarimetric detector
system that is operated on a \mbox{2 m} telescope at the South Pole.  The
resulting polarization map extends $\sim$ 170 pc along the Galactic plane
and $\sim$ 30 pc in Galactic latitude, and thus covers a significant
fraction of the central molecular zone.  Our map shows that this region is
permeated by large-scale toroidal magnetic fields.  We consider our
results together with radio observations that show evidence for poloidal
fields in the Galactic center, and with Faraday rotation observations.  
We compare all of these observations with the predictions of a
magnetodynamic model for the Galactic center that was proposed in order to
explain the Galactic Center Radio Lobe as a magnetically driven gas
outflow.  We conclude that the observations are basically consistent with
the model.

\end{abstract}


\section{Introduction}

Beginning with the discovery of the first examples of Galactic center
non-thermal filaments \citep{yus84}, radio observers have garnered a
steadily growing body of observational evidence pointing toward the
existence of large-scale, ordered magnetic fields in the Galactic center
region, with an overall orientation that is perpendicular to the Galactic
plane \citep{mor98}.  The main tracers of the magnetic field are the
Galactic center non-thermal filaments themselves.  Such filaments have now
been found at a variety of locations throughout the central few hundred pc
\citep{lar00}.  They are generally orthogonal to the Galactic plane, and
radio polarimetry has confirmed that the magnetic fields in these
synchrotron-emitting structures run longitudinally \citep{lan99}.  
Diffuse non-thermal emission is also present, but due to Faraday
rotation/depolarization, polarimetry of this emission does not always
reliably determine the projected field direction.  For the few cases where
it does, the field turns out to again be perpendicular to the Galactic
plane \citep{tsu86}.  This accumulation of radio observations has led to
the hypothesis that most of the volume of the Galactic center is permeated
by a large-scale magnetic field that is poloidal, or perhaps axial.  
(Here we are using a cylindrical coordinate system for the Galaxy, and
poloidal and axial refer respectively to the component in the r-z plane,
and the component along z.)  Direct measurements of the strength of this
large-scale field do not exist, but it has been argued that it must be of
order several mG \citep{mor98}.

Along with this relativistic ionized gas, the Galactic center contains a
concentration of dense molecular gas, extending $\sim$ 400 pc along the
Galactic plane and $\sim$ 50 pc in Galactic latitude \citep{sof98}.  
Far-infrared and submillimeter polarimetric observations of several flux
peaks within this ``central molecular zone'' have revealed a very wide
range of projected magnetic field directions.  Overall, these field
directions are more nearly parallel to the Galactic plane than
perpendicular to it, which has led to the suggestion that the molecular
gas is permeated by a toroidal magnetic field rather than a poloidal or
axial one \citep{mor92, nov00}.

In order to further explore the field in these molecular regions, we have
begun a project called SPARO, that has the goal of making a submillimeter
polarimetric map covering the entire extent of the central molecular zone.  
This requires capability for polarimetry of spatially extended, low
surface-brightness emission.  We achieved this capability by observing at
the South Pole, where the skies are exceptionally transparent and stable
in the submillimeter \citep{lan98}.  In this letter we report on the first
results of the SPARO project.  We describe the observations in \S~2, and
present our polarimetric map in \S~3.  In \S~4 we discuss our results in
the context of a specific model for the Galactic center and its magnetic
field.

\section{Observations}

The observations described here were obtained at the Admundsen-Scott South
Pole station, using the SPARO instrument on the Viper telescope.  SPARO
(the Submillimeter Polarimeter for Antarctic Remote Observations) is a
9-pixel submillimeter array polarimeter incorporating $^3$He-cooled
detectors \citep{dot98, ren01}.  Viper is a \mbox{2-m} diameter Gregorian
off-axis telescope, developed primarily for cosmological observations at
millimeter and submillimeter wavelengths \citep{pet00}.  South Pole
Station is inaccessible between February and October of each year, but it
is during this ``winter'' season that submillimeter observing conditions
are best.  Two of us (G.S.G. and D.P.) remained at South Pole during the
winter season of 2000 to operate the SPARO experiment.  Observations
described here were obtained during April--July 2000.

SPARO's spectral passband is centered at $\lambda_0 = $ 450 \micron, with
fractional bandwidth $\Delta\lambda/\lambda_0=$ 0.10.  The instrument
observes simultaneously at 9 sky positions, with the pixels arranged in a
3 by 3 square pattern.  The pixel-to-pixel separation is 3.5\arcmin, and
the beam FWHM was determined to be 5\arcmin\ $\pm$ 1\arcmin.  The lack of
bright planets observable from the South Pole during the period of the
observations prevented us from determining the beam size more accurately.  
The pointing accuracy was $\pm$ 1\arcmin.  We used two sky reference
positions, separated from the main observing position by +0.5\degr\ and
$-$0.5\degr\ in cross-elevation, respectively.  Moving between main and
reference positions was accomplished by rapidly switching the tilt angle
of Viper's flat mirror (at $\sim$ 3 Hz) while more slowly modulating the
telescope's pointing position (at $\sim$ 0.01 Hz).

The instrumental polarization was determined by observing the center of
the Moon and the peak of Sgr B2 with each pixel.  The Moon was assumed to
be unpolarized near its center, and the peak of Sgr B2 was assumed to have
a polarization of (P = 0.49\%; $\phi$ = 82\degr), obtained by averaging
350 \micron\ polarimetry results for this source \citep{dow98} over the
area of SPARO's beam.  We estimate that the use of 350 \micron\ data to
calibrate our 450 \micron\ observations introduces an error of only $\sim$
0.1\% \citep{dow98, vai01}.

The resulting values for instrumental polarization varied from pixel to
pixel but were all in the range 0.3\%--0.6\%.  Values obtained for a given
pixel using the two independent calibrators typically agreed to within
$\sim$ 0.2\%.  We were also able to test for the ``non-uniform
instrumental polarization'' effect \citep{gon89}, by placing the peak of
Sgr B2 at several different locations within the beam of a pixel and
looking for changes in the measured polarization.  Based on these tests
and calibrations, we are confident that the level of systematic error in
our polarization measurements is $<$ 0.3\%.  This translates into an
uncertainty in polarization angle of $< (9\degr)(P/1.0\%)^{-1}$.

\section{Results}

In Table 1, we present our polarization results with associated
statistical errors for sky positions where we obtained polarization
detections of statistical significance $P/\sigma_P >$ 2.75.  Figure~1
shows these results using bar symbols.  The orientation of each bar gives
the inferred magnetic field direction, that is orthogonal to the E-vector
of the measured polarization, and the length of the bar is proportional to
the degree of polarization.  The bars are superposed on a photometric map
that was also obtained using SPARO.

Clearly visible in the photometric map is the large concentration of
molecular gas that is associated with the innermost few hundred pc of the
Galaxy.  (One degree corresponds to 140 pc for an assumed distance of 8.0
kpc.)  This concentration of gas is asymmetric, with the highest column
density at the position of Sgr B2, displaced toward positive Galactic
longitudes from the location of Sgr A$^*$ by a projected distance of
$\sim$ 100 pc. Our results imply that the magnetic field permeating the
Galactic center molecular gas, when projected onto the plane of the sky,
is for the most part parallel to the Galactic plane.  The simplest
explanation for this alignment of projected field direction with Galactic
plane is that the molecular gas in the Galactic center is threaded by a
large-scale magnetic field having a toroidal (i.e., azimuthal)
configuration.  This has already been suggested based on previous
observations (see \S~1), but the SPARO results provide the strongest
evidence yet obtained for the existence of this toroidal large-scale
field.

\section{Discussion}

Figure 2 contrasts SPARO results with non-thermal filaments (gray scale)
representing evidence for poloidal, or perhaps axial, magnetic fields.
Hereafter, in discussing the field traced by the non-thermal filaments, we
will use only the more general term, poloidal, to describe the inferred
field geometry.  It is clear from Figure~2 that the magnetic field in the
central few hundred pc is neither purely toroidal nor purely poloidal.
Rather, there are regions in which toroidal fields dominate as well as
regions in which poloidal fields dominate.  However, it is not obvious
from this figure how these ``toroidal-dominant'' and ``poloidal-dominant''
regions are arranged with respect to one another in three-dimensional
space.  Furthermore, even when considering the fuller set of observational
evidence discussed in \S~1, one still cannot unambiguously determine this
three-dimensional arrangement.

There does exist a theoretical model for the Galactic center that predicts
separate ``poloidal-dominant'' and ``toroidal-dominant'' regions within
the central few hundred pc.  This is the magnetodynamic model developed by
\citet[hereafter USS85]{uch85}, and further refined by \citet{shi87}. This
model was developed in order to explain the ``Galactic Center Lobe''
(GCL), that is a limb-brightened radio structure with a size of several
hundred pc extending from the plane of the Galaxy up towards positive
Galactic latitudes \citep{sof84}.  In the model of USS85 the GCL
represents a gas outflow that is magnetically driven.  The model consists
of nonsteady axisymmetric magneto-hydrodynamic simulations in which the
field is assumed to be axial at high Galactic latitudes, but acquires a
toroidal component near the Galactic plane due to differential rotation of
the gas to which it is coupled via flux-freezing.  (In the model, the gas
density is higher near the plane.)  The stress of the resultant magnetic
twist is what drives the outflow.

A prediction of this model is that the toroidal component will generally
be more dominant for positions nearer to the Galactic plane \citep{shi87}.  
\citet{yus88} noted that the non-thermal filaments of the Radio Arc (see
Figure~2) cross the Galactic plane without showing any bending, thus
contradicting this prediction.  However, SPARO has now revealed extensive
toroidal-dominant regions near the Galactic plane, as predicted by USS85.  
One way to reconcile Radio Arc observations and SPARO results is to
hypothesize that the filaments of the Radio Arc cross the plane at a
position where molecular gas is largely absent, and that this is what
accounts for the lack of bending.  This is plausible because the in-plane
areal filling factor of the molecular gas is only about 10\%
\citep{bal88}.  Thus the model of USS85 may be correct in an overall sense
while failing to predict the detailed structure of the field because it
ignores the clumpiness of the gas.

Next, we discuss a third probe of the large-scale field in the Galactic
center: Faraday rotation.  This effect is produced in thermal gas lying
along the line-of-sight to the sources of radio synchrotron emission.  
The model of USS85 makes a simple prediction for the large-scale
distribution of the line-of-sight component of the field.  To visualize
this prediction, imagine dividing the Galactic center region into four
sub-regions (quadrants) according to the signs of the Galactic longitude
and Galactic latitude, respectively.  For the purposes of this discussion,
Galactic longitude and latitude are measured with respect to the true
center of the Galaxy at Sgr A$^*$.  USS85 then predict that the
line-of-sight component should have one sign for the (+,+) and ($-$,$-$)
quadrants, and the opposite sign for the (+,$-$) and ($-$,+) quadrants.  
The absolute sign of the line-of-sight component is not predicted by the
model; only the sign changes between quadrants are predicted.  We note
that for every point lying within the three-dimensional volume
corresponding to a given quadrant, the line-of-sight component of the
field has the same sign, according to the model.

USS85 compared this prediction of their model with Faraday rotation
observations that were available at that time.  Specifically, they
considered polarimetry of the ``northern plume'' and ``southern plume'',
two extended synchrotron-emitting regions that appear to be, respectively,
the Galactic northern and Galactic southern extensions of the non-thermal
filaments of the Radio Arc (that are shown in Figure~2).  The two plumes
extend out to Galactic latitudes of about +1\degr\ and $-$1\degr,
respectively.  Faraday rotation measured towards the plumes is believed to
be produced in thermal gas located near to the plumes and associated with
them.  The Faraday rotation measures are predominantly positive for the
northern plume, located in the (+,+) quadrant, and predominantly negative
for the southern plume, located in the (+,$-$) quadrant.  USS85 noted that
such a sign reversal is in agreement with the pattern predicted by their
model.

Since the publication of USS85, a handful of observations of Faraday
rotation of radiation from individual non-thermal filaments have been made
\citep[and references therein]{lan99}.  In these cases, the Faraday
rotation is due to an extended ``Faraday screen'' associated with thermal
gas in the central $\sim$ 200 pc of the Galaxy.  For filaments lying at
positive Galactic longitudes, i.e., in the (+,+) and (+,$-$) quadrants,
the sign of the Faraday rotation observed towards the filaments is in good
agreement with that observed towards the plumes.  Furthermore, we can now
use these more recent measurements to explore the line-of-sight field
towards negative Galactic longitudes, i.e., in the ($-$,+) and ($-$,$-$)
quadrants.  Just two measurements have been published for longitudes
between 0.0\degr\ and $-$1.0\degr.  Specifically, \citet{yus97} observed
negative Faraday rotation measures towards the G359.54+0.18 filament that
lies in the ($-$,+) quadrant, and \citet{gra95} found positive rotation
measures for the filament known as the ``Snake'', that lies in the
($-$,$-$) quadrant.  Combining these data for the ($-$,+) and ($-$,$-$)
quadrants with the pattern already noted by USS85 and given above, namely
positive rotation measures in the (+,+) quadrant and negative rotation
measures in the (+,$-$) quadrant, we find that the predicted and observed
patterns agree precisely.

We have discussed observations made using three different probes of the
magnetic field.  The observations are very sparse, and none of them gives
direct information on the strength of the field.  Field strengths in the
Galactic center have been measured using the Zeeman technique (e.g., see
Yusef-Zadeh et al.\ 1999, and Crutcher et al.\ 1996), but these
observations sample relatively small-scale rather than large-scale fields
so we have not considered them here.  Nevertheless, despite the
limitations of the three kinds of magnetic field observations we have
discussed, they provide significant evidence in favor of the general
picture given by USS85 for the large-scale configuration of the magnetic
field in the Galactic center.

\acknowledgments

For expert technical assistance, crucial help with Antarctic logistics,
and invaluable suggestions, we thank J. Carlstrom, D. Dowell, M. Dragovan,
P. Goldsmith, J. Hanna, A. Harper, R. Hildebrand, R. Hirsch, J. Jaeger, S.
Loverde, P. Malhotra, J. Marshall, H. Moseley, S. Moseley, T. O'Hara, R.
Pernic, S. Platt, J. Sundwall, and J. Wirth.  The SPARO project was funded
by the Center for Astrophysical Research in Antarctica (an NSF Science and
Technology Center; OPP-8920223), by an NSF CAREER Award to G.N.
(OPP-9618319), and by a NASA GSRP award to D.C. (NGT5-88).  We are
grateful to N. E. Kassim, T. N. LaRosa, and D. Pierce-Price for permission
to show their data.

\clearpage

\begin{deluxetable}{r r c c r r}
\tabletypesize{\scriptsize}
\tablecaption{Polarization Results}
\tablewidth{0pt}
\tablehead{
\colhead{$\Delta\alpha$\tablenotemark{a}} &
\colhead{$\Delta\delta$\tablenotemark{a}} &
\colhead{$P(\%)$} &
\colhead{$\sigma_{P}$} &
\colhead{$\phi$\tablenotemark{b}} &
\colhead{$\sigma_{\phi}$}
 }
\startdata

$23\arcmin.1 $ &$44\arcmin.5 $ &$ 2.05 $ &$ 0.61 $ &$ 131.6 $ &$ 9.0 $  \\
$26\arcmin.5 $ &$43\arcmin.5 $ &$ 1.71 $ &$ 0.34 $ &$ 116.7 $ &$ 6.0 $  \\
$25\arcmin.0 $ &$40\arcmin.0 $ &$ 0.67 $ &$ 0.14 $ &$ 95.8 $ &$ 6.5 $  \\ 
$28\arcmin.2 $ &$38\arcmin.7 $ &$ 0.93 $ &$ 0.17 $ &$ 75.6 $ &$ 4.8 $  \\ 
$23\arcmin.7 $ &$36\arcmin.7 $ &$ 0.53 $ &$ 0.06 $ &$ 88.5 $ &$ 3.0 $  \\ 
$26\arcmin.9 $ &$35\arcmin.4 $ &$ 0.96 $ &$ 0.13 $ &$ 78.3 $ &$ 3.4 $  \\ 
$19\arcmin.1 $ &$34\arcmin.8 $ &$ 0.74 $ &$ 0.17 $ &$ 19.7 $ &$ 7.5 $  \\ 
$30\arcmin.1 $ &$34\arcmin.1 $ &$ 1.00 $ &$ 0.33 $ &$ 76.4 $ &$ 8.6 $  \\  
$14\arcmin.9 $ &$32\arcmin.7 $ &$ 1.35 $ &$ 0.43 $ &$ 112.9 $ &$ 9.0 $  \\ 
$25\arcmin.6 $ &$32\arcmin.2 $ &$ 0.68 $ &$ 0.15 $ &$ 89.8 $ &$ 6.3 $  \\
$28\arcmin.8 $ &$30\arcmin.9 $ &$ 1.10 $ &$ 0.34 $ &$ 93.2 $ &$ 8.9 $  \\
$20\arcmin.1 $ &$26\arcmin.8 $ &$ 1.99 $ &$ 0.60 $ &$ 99.1 $ &$ 8.4 $  \\
$15\arcmin.5 $ &$24\arcmin.9 $ &$ 2.55 $ &$ 0.60 $ &$ 100.5 $ &$ 6.6 $  \\
$13\arcmin.9 $ &$20\arcmin.5 $ &$ 2.25 $ &$ 0.76 $ &$ 124.0 $ &$ 9.5 $  \\
$ 8\arcmin.5 $ &$17\arcmin.0 $ &$ 2.13 $ &$ 0.54 $ &$ 121.7 $ &$ 7.2 $  \\
$ 8\arcmin.0 $ &$15\arcmin.3 $ &$ 1.74 $ &$ 0.48 $ &$ 116.9 $ &$ 7.7 $  \\
$ 3\arcmin.8 $ &$14\arcmin.0 $ &$ 1.42 $ &$ 0.25 $ &$ 103.8 $ &$ 4.8 $  \\
$ 7\arcmin.3 $ &$12\arcmin.8 $ &$ 1.08 $ &$ 0.22 $ &$ 119.7 $ &$ 5.8 $  \\
$10\arcmin.5 $ &$11\arcmin.5 $ &$ 1.53 $ &$ 0.34 $ &$ 126.3 $ &$ 6.3 $  \\
$ 2\arcmin.7 $ &$10\arcmin.9 $ &$ 1.34 $ &$ 0.26 $ &$ 114.6 $ &$ 5.4 $  \\
$ 6\arcmin.0 $ &$ 9\arcmin.6 $ &$ 1.02 $ &$ 0.22 $ &$ 112.7 $ &$ 6.4 $  \\
$-5\arcmin.2 $ &$ 5\arcmin.9 $ &$ 2.13 $ &$ 0.74 $ &$ 108.2 $ &$ 9.5 $  \\
$-1\arcmin.5 $ &$ 4\arcmin.8 $ &$ 0.98 $ &$ 0.25 $ &$ 116.7 $ &$ 7.3 $  \\
$ 1\arcmin.8 $ &$ 3\arcmin.5 $ &$ 0.73 $ &$ 0.18 $ &$ 87.6 $ &$ 7.2 $  \\
$ 0\arcmin.5 $ &$ 0\arcmin.2 $ &$ 0.81 $ &$ 0.18 $ &$ 102.5 $ &$ 6.5 $  \\
$ 3\arcmin.7 $ &$-1\arcmin.1 $ &$ 1.09 $ &$ 0.22 $ &$ 103.1 $ &$ 5.9 $  \\
$ 1\arcmin.6 $ &$-1\arcmin.4 $ &$ 2.22 $ &$ 0.49 $ &$ 105.4 $ &$ 6.3 $  \\
$-4\arcmin.5 $ &$-1\arcmin.9 $ &$ 1.99 $ &$ 0.60 $ &$ 117.4 $ &$ 8.4 $  \\
$-0\arcmin.8 $ &$-3\arcmin.0 $ &$ 1.56 $ &$ 0.17 $ &$ 118.9 $ &$ 3.1 $  \\
$ 2\arcmin.4 $ &$-4\arcmin.3 $ &$ 1.57 $ &$ 0.21 $ &$ 110.6 $ &$ 3.8 $  \\
$-7\arcmin.1 $ &$-4\arcmin.6 $ &$ 1.31 $ &$ 0.45 $ &$ 120.1 $ &$ 9.6 $  \\
$-3\arcmin.8 $ &$-5\arcmin.9 $ &$ 0.83 $ &$ 0.27 $ &$ 117.9 $ &$ 9.2 $  \\
$-0\arcmin.6 $ &$-7\arcmin.2 $ &$ 1.53 $ &$ 0.27 $ &$ 120.8 $ &$ 5.1 $  \\
$-8\arcmin.4 $ &$-7\arcmin.8 $ &$ 1.57 $ &$ 0.43 $ &$ 155.9 $ &$ 7.9 $  \\
$-1\arcmin.9 $ &$-10\arcmin.5 $ &$ 2.00 $ &$ 0.31 $ &$ 116.2 $ &$ 4.5 $  \\
$-6\arcmin.4 $ &$-12\arcmin.4 $ &$ 2.31 $ &$ 0.40 $ &$ 122.8 $ &$ 4.9 $  \\
$-3\arcmin.2 $ &$-13\arcmin.7 $ &$ 1.48 $ &$ 0.36 $ &$ 119.5 $ &$ 6.7 $  \\ 
\hline
\enddata

\tablenotetext{a}{Offsets in Right Ascension and Declination are
measured relative to the position of Sgr A$^*$.}

\tablenotetext{b}{$\phi$ is the angle of the E-vector of the polarized
radiation, measured in degrees from north-south, increasing
counterclockwise.}

\end{deluxetable}

\clearpage


\begin{figure}
\epsscale{0.9}
\plotone{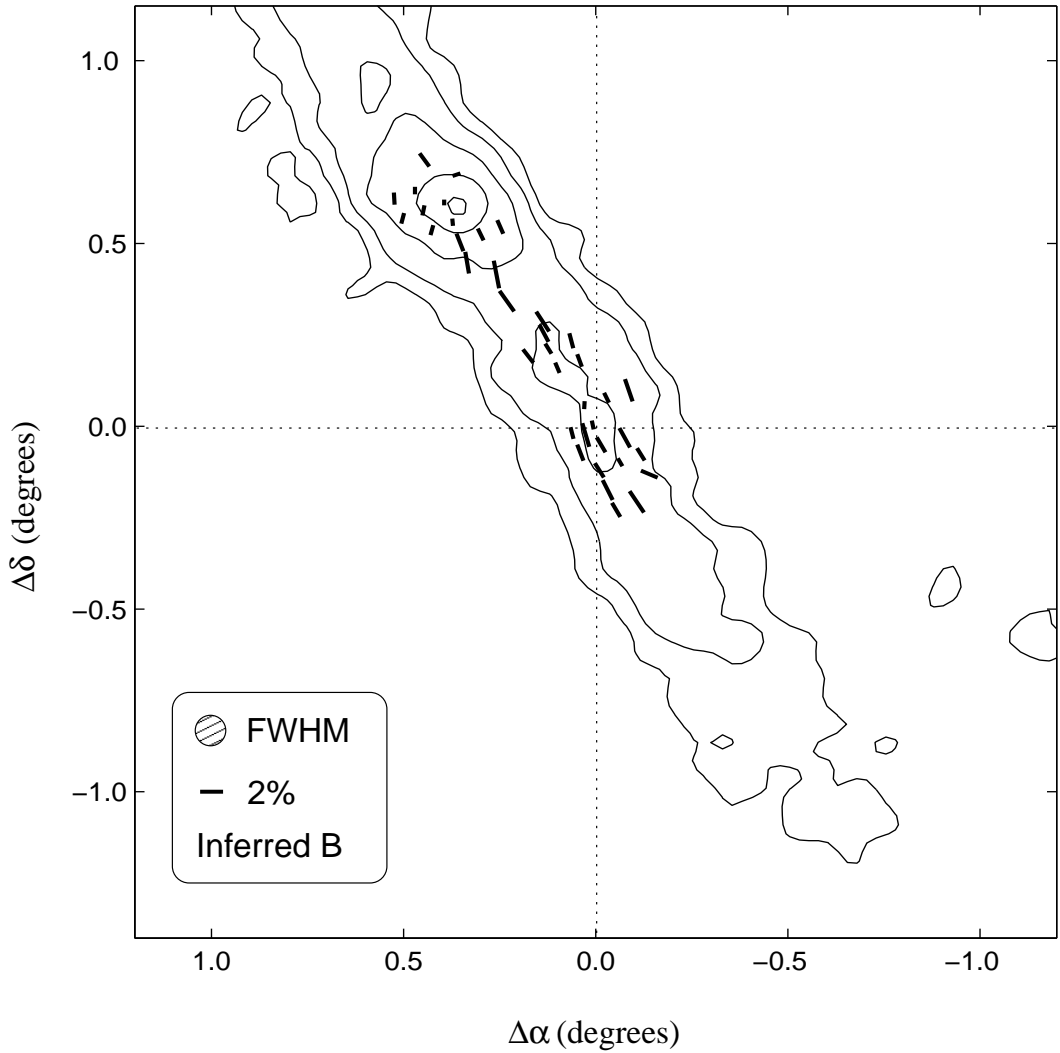}

\caption{Results of 450 \micron\ polarimetry (bars) and photometry
(contours) of the Galactic center, obtained using SPARO.  The distribution
of 450 \micron\ flux closely follows the Galactic plane, that lies at a
position angle of +31\degr.  Coordinate offsets are measured with respect
to the location of Sgr A$^*$ (that lies at the intersection of the
horizontal and vertical dotted lines).  Each bar is drawn parallel to the
inferred magnetic field direction (i.e. perpendicular to the E-vector of
the measured submillimeter polarization), and the length of the bar
indicates the measured degree of polarization (see key at bottom left).  
Contours are drawn at 0.075, 0.15, 0.30, 0.60, and 0.95 times the peak
flux, which is located at the position of Sgr B2.  For clarity, negative
contours are not shown.  The reference beam offsets were the same for
polarimetry and photometry and are given in \S~2.  The 5\arcmin\ beam of
SPARO is shown in the key.  Positive Galactic latitudes lie towards the
upper right of the figure, and positive Galactic longitudes lie towards
upper left.}

\end{figure}

\begin{figure}
\epsscale{0.9}
\plotone{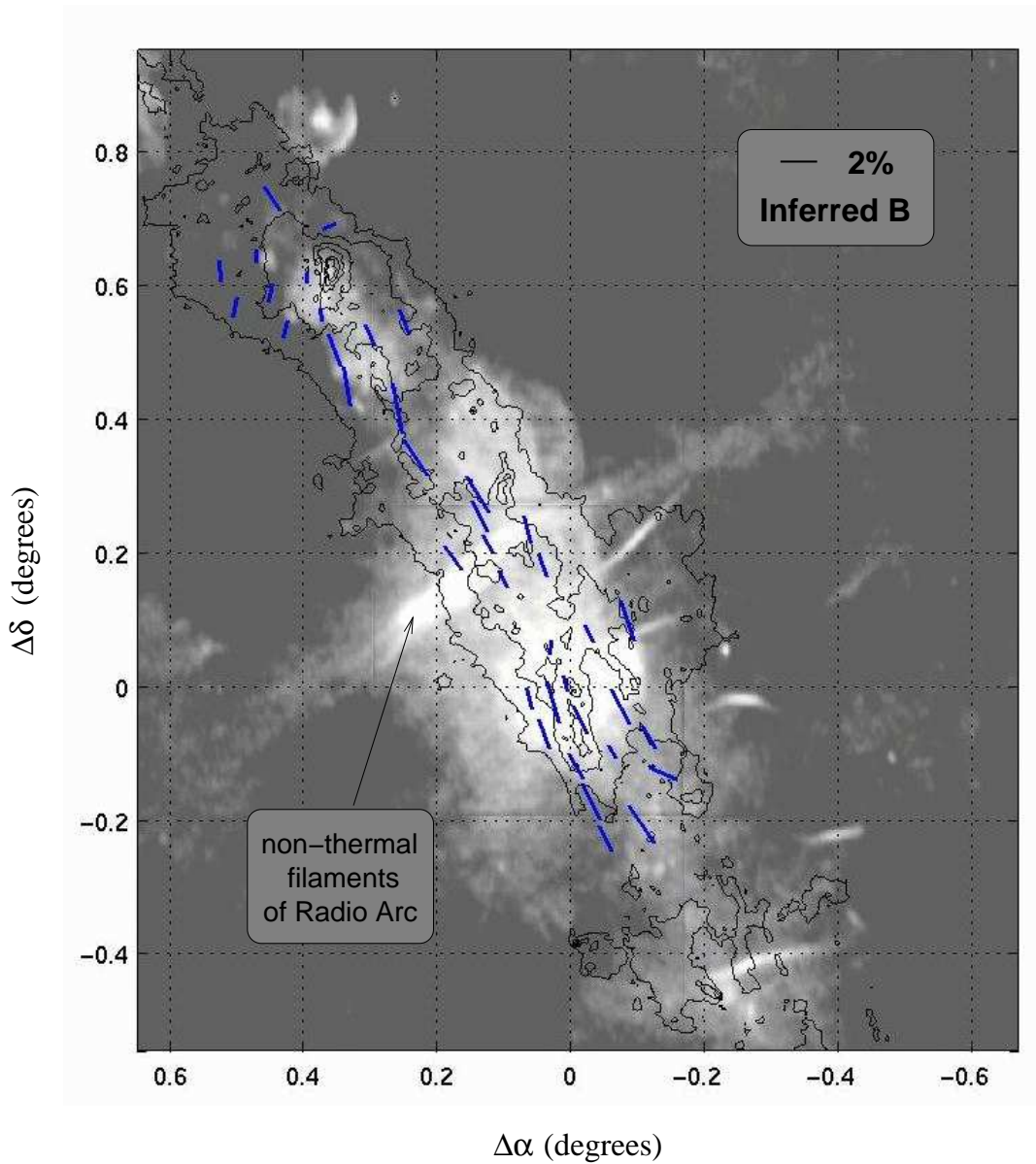}

\caption{450 \micron\ polarization measurements (bars) shown together with
90 cm radio continuum image (gray scale, LaRosa et al.\ 2000), and 850
\micron\ continuum emission (contours, Pierce-Price et al. 2000).  As in
Fig. 1, the orientation of each bar is parallel to the inferred magnetic
field direction (i.e., orthogonal to the measured direction of
polarization) and its length is proportional to the degree of
polarization.  The radio continuum image shows about six locations where
non-thermal filaments can be seen.  These non-thermal filaments trace
magnetic fields in hot ionized regions (see \S~1).  The gray scale image
is logarithmically scaled, and the contours of 850~\micron\ emission are
also logarithmic.  Coordinate offsets are measured with respect to the
position of Sgr A$^*$.  The location of the brightest bundle of
non-thermal filaments (referred to as the non-thermal filaments of the
Radio Arc) is indicated in the figure.}

\end{figure}

\end{document}